\documentclass{aa}  

\usepackage{graphicx}
\usepackage{txfonts}
\usepackage{color}
\usepackage{stfloats}

\newcommand{\tablenotea}[1]{\parbox{18.0cm}{\indent \footnotesize{#1}}}

\newcommand{\cpl}{Chem. Phys. Lett.}
\newcommand{\chemrev}{Chem. Rev.}
\newcommand{\cflame}{Combust. Flame}
\newcommand{\ieeetap}{IEEE Trans. Antennas Propag.}
\newcommand{\ijms}{Int. J. Mass Spectrom.}
\newcommand{\jacs}{J. Am. Chem. Soc.}
\newcommand{\jmst}{J. Mol. Struct.}
\newcommand{\jpc}{J. Phys. Chem.}
\newcommand{\jpca}{J. Phys. Chem. A}
\newcommand{\jpcrd}{J. Phys. Chem. Ref. Data}
\newcommand{\jtki}{Jurnal Teknik Kimia Indonesia}
\newcommand{\natastro}{Nat. Astron.}
\newcommand{\pccp}{Phys. Chem. Chem. Phys.}
\newcommand{\pci}{Proc. Combust. Inst.}
\newcommand{\science}{Science}
\newcommand{\zpc}{Z. Phys. Chem.}

\begin{document}

\title{Discovery of the propargyl radical (CH$_2$CCH) in \mbox{TMC-1}:\\ one of the most abundant radicals ever found and a key species for cyclization to benzene in cold dark clouds\thanks{Based on observations carried out with the Yebes 40m telescope (projects 19A003, 20A014, and 20D023). The 40m radiotelescope at Yebes Observatory is operated by the Spanish Geographic Institute (IGN, Ministerio de Transportes, Movilidad y Agenda Urbana).}}

\titlerunning{Detection of CH$_2$CCH in \mbox{TMC-1}}
\authorrunning{Ag\'undez et al.}

\author{M.~Ag\'undez\inst{1}, C. Cabezas\inst{1}, B.~Tercero\inst{2,3}, N.~Marcelino\inst{1}, J.~D.~Gallego\inst{3}, P.~de~Vicente\inst{3}, \and J.~Cernicharo\inst{1}}

\institute{
Instituto de F\'isica Fundamental, CSIC, Calle Serrano 123, E-28006 Madrid, Spain\\ \email{marcelino.agundez@csic.es, jose.cernicharo@csic.es} \and
Observatorio Astron\'omico Nacional, IGN, Calle Alfonso XII 3, E-28014 Madrid, Spain \and
Observatorio de Yebes, IGN, Cerro de la Palera s/n, E-19141 Yebes, Guadalajara, Spain
}

\date{Received; accepted}

 
\abstract
{We present the first identification in interstellar space of the propargyl radical (CH$_2$CCH). This species was observed in the cold dark cloud \mbox{TMC-1} using the Yebes 40m telescope. The six strongest hyperfine components of the 2$_{0,2}$-1$_{0,1}$ rotational transition, lying at 37.46~GHz, were detected with signal-to-noise ratios in the range 4.6-12.3\,$\sigma$. We derive a column density of $8.7 \times 10^{13}$~cm$^{-2}$ for CH$_2$CCH, which translates to a fractional abundance relative to H$_2$ of $8.7 \times 10^{-9}$. This radical has a similar abundance to methyl acetylene, with an abundance ratio CH$_2$CCH/CH$_3$CCH close to one. The propargyl radical is thus one of the most abundant radicals detected in \mbox{TMC-1}, and it is probably the most abundant organic radical with a certain chemical complexity ever found in a cold dark cloud. We constructed a gas-phase chemical model and find calculated abundances that agree with, or fall two orders of magnitude below, the observed value depending on the poorly constrained low-temperature reactivity of CH$_2$CCH with neutral atoms. According to the chemical model, the propargyl radical is essentially formed by the C + C$_2$H$_4$ reaction and by the dissociative recombination of C$_3$H$_n$$^+$ ions with $n$ = 4-6. The propargyl radical is believed to control the synthesis of the first aromatic ring in combustion processes, and it probably plays a key role in the synthesis of large organic molecules and cyclization processes to benzene in cold dark clouds.}


\maketitle

\section{Introduction}

Cold dark clouds like \mbox{TMC-1} have revealed as extraordinary chemical laboratories, able to synthesize in situ a great variety of molecules. Most detected species are neutral. Some cations have been detected, mostly protonated forms of closed-shell abundant molecules (e.g., \citealt{Agundez2015b,Marcelino2020,Cernicharo2020c,Cernicharo2021a,Cernicharo2021b}), and a few hydrocarbon and nitrile anions have been also observed (e.g., \citealt{Cernicharo2020a}), but the vast majority of species detected are electrically neutral. In general, molecular ions are observed with low abundances, below 10$^{-10}$ relative to H$_2$, due to their high reactivity.

A large fraction of the neutral species observed in cold dark clouds, and the most abundant ones, are closed-shell molecules. Among them, the long known unsaturated carbon chains stand out as the most prevalent type of molecules \citep{Agundez2013}. However, in recent times it has been found  that cold dark clouds contain also a variety of organic molecules of increasing complexity, going from the nearly saturated propylene \citep{Marcelino2007}, various isomers of the partially saturated molecules C$_4$H$_4$, C$_5$H$_4$, C$_4$H$_3$N, and C$_5$H$_3$N \citep{Cernicharo2021c,Cernicharo2021d,Marcelino2021,McGuire2020,Lee2021}, the five-membered ring C$_5$H$_5$CN \citep{McCarthy2020}, and the aromatic ring C$_6$H$_5$CN \citep{McGuire2018,Burkhardt2021}. These detections reveal that there are chemical processes not yet well characterized that are able to synthesize large complex organic molecules under very cold conditions.

About two thirds of the neutral species detected in these cold environments are open-shell radicals. With the exception of a few small radicals like OH, CH, C$_2$H, C$_4$H, and NO, observed radicals have low abundances because, as ions, they are highly reactive species (see, e.g., \citealt{Agundez2013}). In addition, they tend to suffer from spectral dilution due to line splitting resulting from the coupling of the electron spin with the rotation, and also often with the spin of nuclei. These facts complicate the detection of radicals in cold clouds. Recent examples of detections of radicals are HCCO, HCS, and NCO \citep{Agundez2015a,Agundez2018,Marcelino2018}. However, the chemical routes to form molecules of increasing complexity are made of reactions involving reactive species like radicals and ions. Detecting them is therefore of paramount importance to unveil the synthetic pathways postulated by chemical models. Here we report the detection of the propargyl radical (CH$_2$CCH) in \mbox{TMC-1}. This species is one of the most abundant radicals ever found in cold dark clouds, and it is a potential key intermediate in the formation of complex organic molecules like aromatic rings. 

\section{Observations}

The propargyl radical (CH$_2$CCH) is an asymmetric rotor, with a planar geometry, C$_{2v}$ symmetry, and a ground electronic state $^2B_1$. Calculations indicate that this radical has a very small electric dipole moment, 0.14~D according to \cite{Botschwina1995} and 0.115~D according to \cite{Woon2009}, something that was confirmed experimentally by measuring it to be 0.150~D \citep{Kupper2002}. In spite of this small dipole moment, \cite{Tanaka1997} were able to measure in the laboratory the fine and hyperfine structure of various rotational transitions lying at 18.7~GHz and in the 37-38~GHz range, with frequency uncertainties of a few kHz.

Based on their laboratory measurements, \cite{Tanaka1997} searched for the 2$_{0,2}$-1$_{0,1}$ transition of CH$_2$CCH, lying at 37.46~GHz, toward \mbox{TMC-1} using the Nobeyama 45m telescope. They did not detect the radical after an integration time of 6~h in which they reached a noise level of 12~mK per 37~kHz channel. Our line survey of \mbox{TMC-1} carried out with Yebes 40m telescope has a much higher sensitivity, with a $T_A^*$ rms noise level of 0.30~mK per 38.15~kHz channel in the region around 37.46~GHz, which has enabled a robust detection of the propargyl radical.

\begin{figure*}
\centering
\includegraphics[angle=0,width=0.90\textwidth]{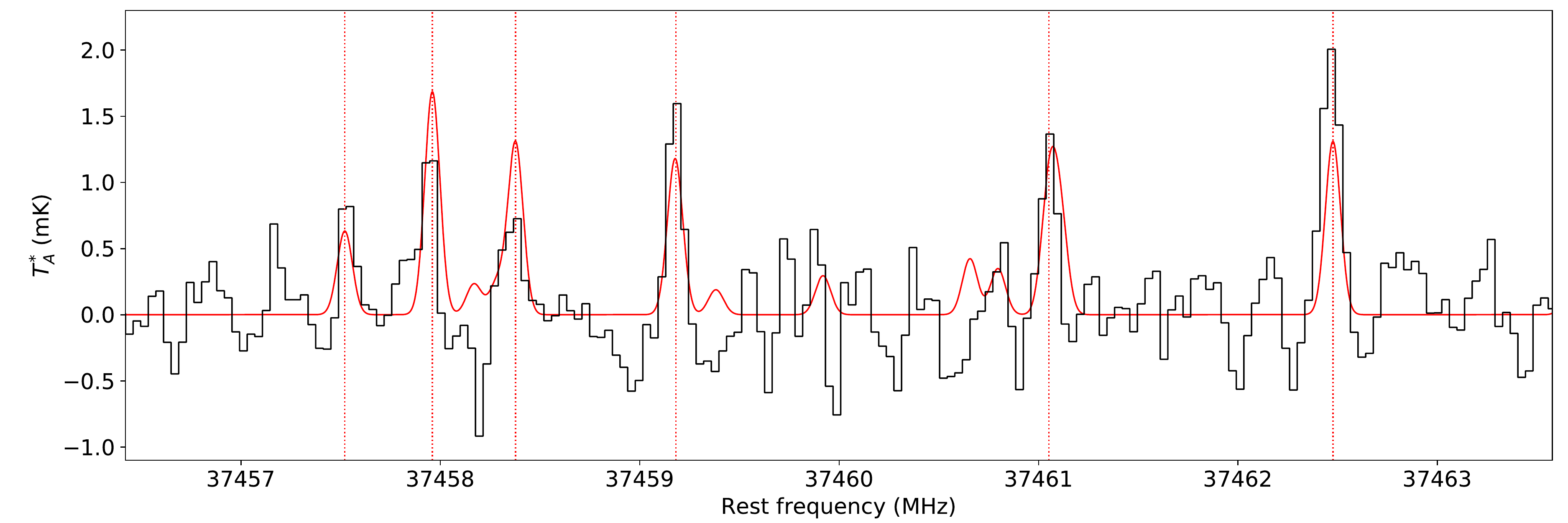}
\caption{Spectrum of \mbox{TMC-1} taken with the Yebes 40m telescope around 37.46~GHz (black histogram). The frequencies of the six most intense hyperfine components of the 2$_{0,2}$-1$_{0,1}$ transition of ortho CH$_2$CCH are indicated by red dotted vertical lines. Transition quantum numbers, frequencies, and derived line parameters are given in Table~\ref{table:lines}. The synthetic spectrum (red line) was computed for a column density of ortho CH$_2$CCH of $6.5\times10^{13}$ cm$^{-2}$, a rotational temperature of 10 K, an emission size of 40$''$ of radius, and a linewidth of 0.72 km s$^{-1}$ (see text).} \label{fig:lines}
\end{figure*}

The data in which the detection of CH$_2$CCH is based is part of a line survey of \mbox{TMC-1} in the Q band carried out with the Yebes 40m telescope. We used the cryogenic receiver for the Q band, built within the Nanocosmos project\,\footnote{\texttt{https://nanocosmos.iff.csic.es}}, which covers the 31.0-50.4~GHz frequency range with horizontal and vertical polarizations. Receiver temperatures vary from 17~K at 32~GHz to 25~K at 50~GHz. The spectrometers are FFTS, which cover a bandwidth of $8\times2.5$~GHz in each polarization with a spectral resolution of 38.15~kHz. The system is described in detail by \cite{Tercero2021}. The line survey was carried out during various observing sessions and several results have been already published. Results like the detection of the negative ions C$_3$N$^-$ and C$_5$N$^-$ \citep{Cernicharo2020a}, the discoveries of HC$_4$NC \citep{Cernicharo2020b}, HC$_3$O$^+$ \citep{Cernicharo2020c}, and HC$_5$NH$^+$ \citep{Marcelino2020} were based on the data taken during November 2019 and February 2020. An additional observing run was carried out in October 2020, which allowed to improve the sensitivity and resulted in the detection of HDCCN \citep{Cabezas2021}, HC$_3$S$^+$ \citep{Cernicharo2021a}, and CH$_3$CO$^+$ \citep{Cernicharo2021b}, and the observational study of C$_4$H$_3$N isomers \citep{Marcelino2021}. A final observing run was performed in December 2020 and January 2021, resulting in the discovery of CH$_2$CHCCH \citep{Cernicharo2021c} and CH$_2$CCHCCH \citep{Cernicharo2021d}. All observations were carried out using the frequency switching technique, with a frequency throw of 10 MHz during the two first observing runs and 8 MHz in the later ones. The intensity scale, the antenna temperature $T_A^*$, was calibrated using two absorbers at different temperatures and the atmospheric transmission model {\small ATM} \citep{Cernicharo1985,Pardo2001}. The uncertainty in $T_A^*$ is estimated to be around 10\,\%. To convert to main beam brightness temperature ($T_{\rm mb}$) one has to divide $T_A^*$ by $B_{\rm eff}$/$F_{\rm eff}$. The parameters of the Yebes 40m antenna\,\footnote{\texttt{http://rt40m.oan.es/rt40m\_en.php}} at the frequency of interest here, 37.46~GHz, are $B_{\rm eff}$ = 0.56 and $F_{\rm eff}$ = 0.97, while the half power beam width (HPBW) is 47.8$''$. All data have been reduced with the program {\small CLASS} of the {\small GILDAS} software package\,\footnote{\texttt{http://www.iram.fr/IRAMFR/GILDAS}}.

\section{Results}

\begin{table*}
\small
\caption{Observed line parameters of CH$_2$CCH in \mbox{TMC-1}.}
\label{table:lines}
\centering
\begin{tabular}{lccccccc}
\hline \hline
\multicolumn{1}{c}{Transition\,$^a$} & \multicolumn{1}{c}{$\nu_{\rm calc}$\,$^b$} & \multicolumn{1}{c}{$A_{ul}$\,$^c$} & \multicolumn{1}{c}{$T_A^*$ peak} & \multicolumn{1}{c}{$\Delta v$\,$^d$}      & \multicolumn{1}{c}{$V_{\rm LSR}$}      & \multicolumn{1}{c}{$\int T_A^* dv$} & \multicolumn{1}{c}{SNR\,$^e$} \\
\multicolumn{1}{c}{} & \multicolumn{1}{c}{(MHz)}        & \multicolumn{1}{c}{(s$^{-1}$)}        & \multicolumn{1}{c}{(mK)}                   & \multicolumn{1}{c}{(km s$^{-1}$)}  & \multicolumn{1}{c}{(km s$^{-1}$)}  & \multicolumn{1}{c}{(mK km s$^{-1}$)} & \multicolumn{1}{c}{($\sigma$)} \\
\hline
2$_{0,2}$-1$_{0,1}$ \hspace{0.1cm} $J$ = 3/2-1/2 \hspace{0.1cm} $F_1$ = 2-1 \hspace{0.1cm} $F$ = 2-1 & 37457.521 & $3.72 \times 10^{-9}$ & $0.96 \pm 0.30$ & $0.61 \pm 0.26$ & $5.57 \pm 0.13$ & $0.62 \pm 0.24$ & 4.8 \\ 
2$_{0,2}$-1$_{0,1}$ \hspace{0.1cm} $J$ = 5/2-3/2 \hspace{0.1cm} $F_1$ = 3-2 \hspace{0.1cm} $F$ = 4-3 & 37457.960 & $5.51 \times 10^{-9}$ & $1.32 \pm 0.30$ & $0.66 \pm 0.22$ & $5.84 \pm 0.09$ & $0.93 \pm 0.24$ & 6.9 \\ 
2$_{0,2}$-1$_{0,1}$ \hspace{0.1cm} $J$ = 5/2-3/2 \hspace{0.1cm} $F_1$ = 2-1 \hspace{0.1cm} $F$ = 3-2 & 37458.377 & $5.45 \times 10^{-9}$ & $0.75 \pm 0.30$ & $0.89 \pm 0.43$ & $5.80 \pm 0.23$ & $0.71 \pm 0.34$ & 4.6 \\ 
2$_{0,2}$-1$_{0,1}$ \hspace{0.1cm} $J$ = 5/2-3/2 \hspace{0.1cm} $F_1$ = 3-2 \hspace{0.1cm} $F$ = 3-2 & 37459.181 & $3.71 \times 10^{-9}$ & $1.73 \pm 0.30$ & $0.64 \pm 0.12$ & $5.72 \pm 0.06$ & $1.17 \pm 0.21$ & 8.8 \\ 
2$_{0,2}$-1$_{0,1}$ \hspace{0.1cm} $J$ = 5/2-3/2 \hspace{0.1cm} $F_1$ = 2-1 \hspace{0.1cm} $F$ = 2-1 & 37461.052 & $5.45 \times 10^{-9}$ & $1.39 \pm 0.30$ & $0.68 \pm 0.15$ & $5.68 \pm 0.07$ & $1.01 \pm 0.20$ & 7.3 \\ 
2$_{0,2}$-1$_{0,1}$ \hspace{0.1cm} $J$ = 3/2-1/2 \hspace{0.1cm} $F_1$ = 2-1 \hspace{0.1cm} $F$ = 3-2 & 37462.476 & $5.50 \times 10^{-9}$ & $2.07 \pm 0.30$ & $0.86 \pm 0.10$ & $5.78 \pm 0.05$ & $1.90 \pm 0.20$ & 12.3 \\ 
\hline
\end{tabular}
\tablenotea{\\
The line parameters $T_A^*$ peak, $\Delta v$, $V_{\rm LSR}$, and $\int T_A^* dv$ and the associated errors are derived from a Gaussian fit to each line profile. $^a$ Quantum numbers from the coupling scheme of \cite{Tanaka1997}. $^b$ Calculated frequencies $\nu_{\rm calc}$ from the {\small CDMS} entry. $^c$ $A_{ul}$ is the Einstein coefficient of spontaneous emission. $^d$ $\Delta v$ is the full width at half maximum (FWHM). $^e$ Signal-to-noise ratio is computed for a spectral resolution equal to the linewidth. That is, \mbox{SNR = [S]/[N]}, where the signal is evaluated from the line area as \mbox{[S] = [$\int T_A^* dv$ / $\Delta v$]}, while the noise is evaluated as \mbox{[N] = [rms $\times$ $\sqrt{\delta \nu / (\Delta v \times \nu_{\rm calc} / c)}$]}, where we used the inverse square-root dependence of noise on spectral resolution, as given by the radiometer equation. In the expressions of [S] and [N], rms is 0.30~mK, $\delta \nu$ is the spectral resolution (0.03815 MHz), $c$ is the speed of light in km s$^{-1}$, and the rest of parameters are given in the table with the appropriate units.
}
\end{table*}

The spectrum of \mbox{TMC-1} at 37.46~GHz has a $T_A^*$ rms noise level of 0.30~mK and show various emission lines which can be assigned to the six strongest hyperfine components of the 2$_{0,2}$-1$_{0,1}$ transition of ortho CH$_2$CCH (see Fig.~\ref{fig:lines}). The energy levels, transition frequencies, and line strengths of CH$_2$CCH were taken from the Cologne Database for Molecular Spectroscopy ({\small CDMS}; \citealt{Muller2005})\,\footnote{\texttt{https://cdms.astro.uni-koeln.de/}}, which is based on a fit to the laboratory frequencies measured by \cite{Tanaka1997}, and is implemented in {\small MADEX} \citep{Cernicharo2012}. We considered separately the ortho and para species and adopted a dipole moment of 0.150~D, as measured by \cite{Kupper2002}.

The six observed lines are precisely centered at the calculated frequencies from the {\small CDMS}, adopting a systemic velocity of 5.83 km s$^{-1}$ for \mbox{TMC-1} \citep{Cernicharo2020b}. The values of $V_{\rm LSR}$ derived are very close to this value (see Table~\ref{table:lines}), with deviations in frequency of 32~kHz for one line and $<20$~kHz for the rest, which is within the uncertainty given by the spectral resolution of 38.15~kHz and the error in the Gaussian fit. The six lines are also detected at a significant level, between $4.6\,\sigma$ and $12.3\,\sigma$ (see Table~\ref{table:lines}). We therefore consider that the detection of CH$_2$CCH in \mbox{TMC-1} is robust.

To further support the detection of the propargyl radical in \mbox{TMC-1} we calculated a synthetic spectrum. The observed lines have similar upper level energies and thus it is not possible to derive a rotational temperature. Nevertheless, given the low dipole moment of CH$_2$CCH (0.150~D), which results in low critical densities (probably as low as a few 10$^2$ cm$^{-3}$), we can safely assume that rotational levels are thermalized and the rotational temperature is equal to the kinetic temperature of $\sim$\,10~K \citep{Feher2016}. To compute the synthetic spectrum we adopted a full width at half maximum (FWHM) of 0.72~km s$^{-1}$, which is the arithmetic mean of the values derived from the six observed lines (see Table~\ref{table:lines}), and assumed that the emission from CH$_2$CCH is distributed in the sky as a circle with a radius of 40$''$, as observed for various hydrocarbons in \mbox{TMC-1} \citep{Fosse2001}. The observed intensities are reproduced with a column density of ortho CH$_2$CCH of $6.5\times10^{13}$ cm$^{-2}$. As shown in Fig.~\ref{fig:lines}, the computed relative and absolute intensities of the six hyperfine components agree well with the observed ones. It is also seen that weaker components are not detected because the calculated intensities are within the noise of the spectrum. That is, there are no missing hyperfine components of the 2$_{0,2}$-1$_{0,1}$ transition.

There are two rotational transitions of para CH$_2$CCH covered by our Q band line survey of \mbox{TMC-1}, the 2$_{1,2}$-1$_{1,1}$ and the 2$_{1,1}$-1$_{1,0}$, lying a 37.2~GHz and 37.8~GHz, respectively. The ground state of the para species is located 14.3~K above the ground state of the ortho species. Treating ortho and para as separate species and assuming an ortho-to-para ratio equal to the statistical value of 3, the most intense component of these two transitions is calculated with $T_A^*$ = 0.5~mK, which is within the noise in these spectral regions. A deeper integration that reduces the noise level should allow to detect these lines. Therefore, the non detection of these two transitions of para CH$_2$CCH is consistent with the detection of the transition of ortho CH$_2$CCH shown in Fig.~\ref{fig:lines}. Assuming an ortho-to-para ratio of 3, the column density derived for CH$_2$CCH (including ortho and para species) in \mbox{TMC-1} is $8.7 \times 10^{13}$~cm$^{-2}$. Adopting a column density of H$_2$ of 10$^{22}$~cm$^{-2}$ \citep{Cernicharo1987}, the fractional abundance of CH$_2$CCH relative to H$_2$ is $8.7 \times 10^{-9}$.

\section{Discussion}

It is remarkable that the propargyl radical is one of the most abundant radicals ever found in \mbox{TMC-1}. Only a few simple radicals like OH, CH, C$_4$H, and NO have been detected with abundances above that derived for CH$_2$CCH (see, e.g., \citealt{Agundez2013}). Moreover, CH$_2$CCH has an abundance similar to its closed-shell counterpart CH$_3$CCH, which has a column density of (1.1-1.3)\,$\times 10^{14}$~cm$^{-2}$ \citep{Gratier2016,Cabezas2021}. Therefore, the abundance ratio CH$_2$CCH/CH$_3$CCH is close to one. This fact is unusual since in \mbox{TMC-1}, in particular, and in cold dark clouds, in general, a radical resulting from removing one hydrogen atom from a closed-shell molecule is usually less abundant than the corresponding closed-shell molecule. The only other example of a radical being more abundant than the corresponding closed-shell molecule is the cyanomethyl radical, in which case the abundance ratio CH$_2$CN/CH$_3$CN in \mbox{TMC-1} is in the range 3-9 \citep{Gratier2016,Cabezas2021}. The large abundance of CH$_2$CCH and the fact that it is as abundant as, but more reactive than, CH$_3$CCH makes this radical a very interesting intermediate in the build up of chemical complexity in cold dark clouds like \mbox{TMC-1}.

To get insight into the chemical implications of the detection of the propargyl radical in \mbox{TMC-1} we carried out chemical model calculations. We are interested in various aspects: whether the high abundance derived for CH$_2$CCH can be accounted for by a gas-phase chemical model, which are the main formation and destruction pathways of CH$_2$CCH, and which kind of chemical routes are opened by the presence of this abundant radical. We adopt typical parameters of cold dark clouds, i.e., a gas kinetic temperature of 10~K, a volume density of H$_2$ of $2 \times 10^4$~cm$^{-3}$, a visual extinction of 30~mag, a cosmic-ray ionization rate of H$_2$ of $1.3 \times 10^{-17}$~s$^{-1}$, and the set of "low-metal" elemental abundances (see, e.g., \citealt{Agundez2013}). We use the gas-phase chemical network {\small RATE12} from the {\small UMIST} database \citep{McElroy2013}, where we have revised the reactions involved in the formation and destruction of CH$_2$CCH and the related molecules CH$_3$CCH and CH$_2$CCH$_2$ (see Table~\ref{table:reactions}).

\begin{figure}
\centering
\includegraphics[angle=0,width=\columnwidth]{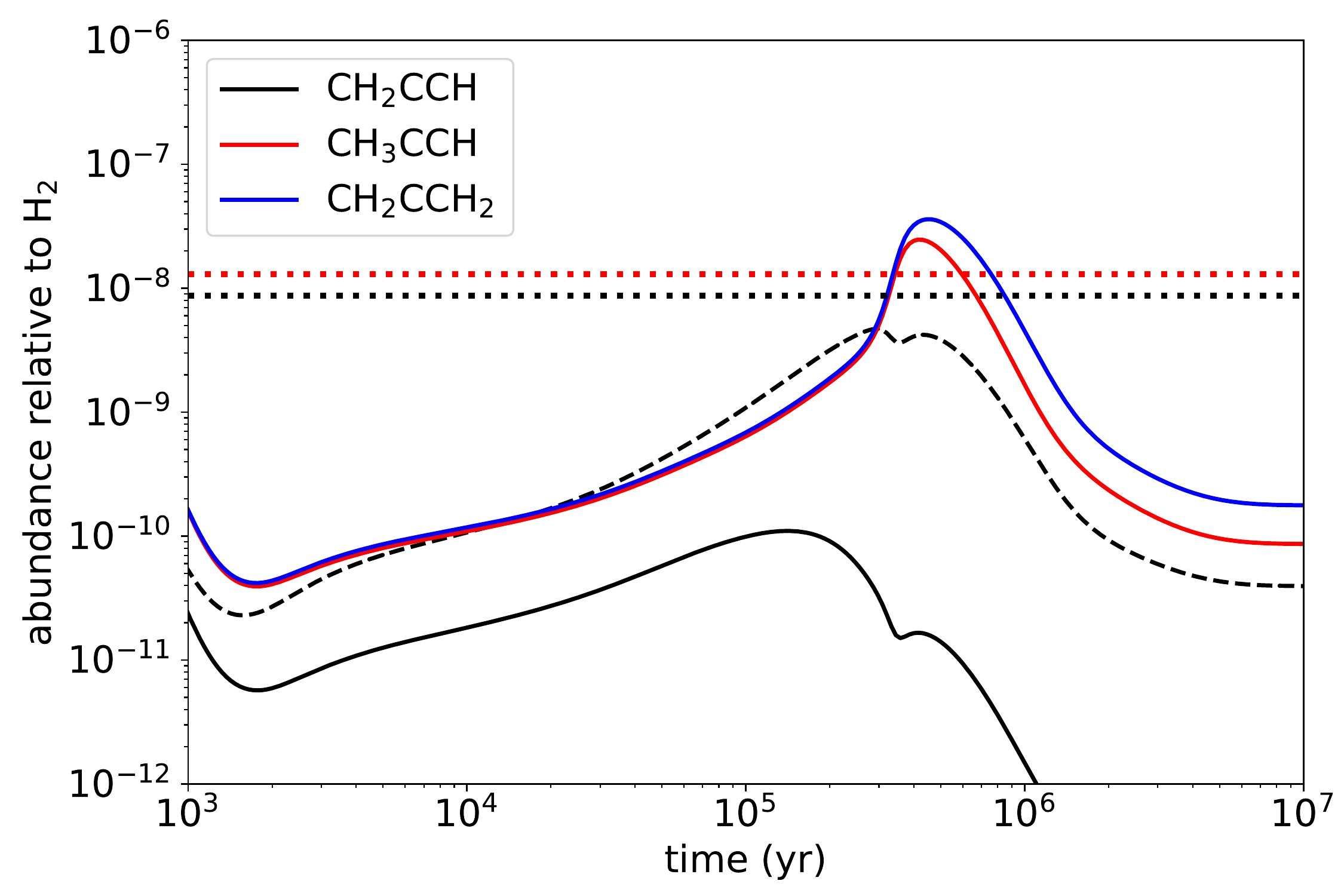}
\caption{Calculated fractional abundances of CH$_2$CCH and related molecules as a function of time. The black dashed line corresponds to the calculated abundance of CH$_2$CCH when the reactions of this radical with O and N atoms are removed. The abundances observed in \mbox{TMC-1} are indicated by dotted horizontal lines.} \label{fig:abun}
\end{figure}

The abundance calculated for the propargyl radical is shown in Fig.~\ref{fig:abun}, together with those of methyl acetylene and allene. It is seen that the calculated abundances of CH$_3$CCH and CH$_2$CCH$_2$ are very similar at any time, which points to allene being as abundant as methyl acetylene in cold dark clouds. Moreover, the peak abundance of CH$_3$CCH is close to the observed value. In the case of the radical CH$_2$CCH, the calculated abundance lies well below the observed value and the CH$_2$CCH/CH$_3$CCH abundance ratio falls well below one, in contrast with observations. To better understand the behavior of the calculated abundances it is useful to examine which are the main formation and destruction processes of each species in the chemical model. The propargyl radical is mostly formed through the neutral-neutral reaction
\begin{equation} \label{reac:c+c2h4}
\rm C + C_2H_4 \rightarrow \rm CH_2CCH + H,
\end{equation}
which has been measured to be rapid down to 15~K \citep{Chastaing1999} and yields propargyl radical as main product, as indicated by extensive experimental and theoretical evidence \citep{Kaiser1996,Bergeat2001,Geppert2003,Chin2012,Mandal2018}. A second formation pathway, which is also the main route to the two C$_3$H$_4$ isomers, is provided by the dissociative recombination of C$_3$H$_n$$^+$ ions with electrons
\begin{equation} \label{reac:c3hn+}
\rm C_3H_{\textit n}^{+} + e^- \rightarrow \rm products,
\end{equation}
with $n$ = 4-7, since ions with $n > 7$ are not included in the chemical network. Information on the product branching ratios is available to different degrees of detail for the dissociative recombination of the different ions C$_3$H$_n$$^+$ \citep{Angelova2004,Geppert2004,Ehlerding2003,Larsson2005}. For example, it is known that the CH$_2$CCH radical is the main product in the dissociative recombination of C$_3$H$_4$$^+$ \citep{Geppert2004}, while it is not formed at all during the dissociative recombination of C$_3$H$_7$$^+$ \citep{Ehlerding2003,Larsson2005}. Thus, the ions C$_3$H$_n$$^+$ with $n = 4$-6 contribute to the formation of CH$_2$CCH, while the closed-shell hydrocarbons CH$_3$CCH and CH$_2$CCH$_2$ are essentially formed upon dissociative recombination of C$_3$H$_n$$^+$ ions with $n = 5$-7 \citep{Ehlerding2003,Angelova2004,Larsson2005}. The synthesis of the ions C$_3$H$_n$$^+$ occurs through a series of reactions involving cations (see, e.g., \citealt{Herbst1989}).

Reactions with neutral atoms are the main destruction process for the propargyl radical and the closed-shell molecules methyl acetylene and allene. There is experimental evidence that C atoms react fast with CH$_3$CCH and CH$_2$CCH$_2$ \citep{Loison2004}, although these hydrocarbons do not react rapidly with either O or N atoms at low temperatures due to the presence of activation barriers \citep{Adusei1996,Michael1977}. Information on the reactivity of CH$_2$CCH with neutral atoms is more limited. The propargyl radical reacts fast with O atoms at room and high temperature \citep{Slagle1991}, although it is not known whether this behavior is maintained at very low temperatures. The radical is also thought to react fast with C and N atoms \citep{Smith2004,Loison2017}, although there is no experimental or theoretical evidence. The reaction between CH$_2$CCH and H atoms has an activation barrier to produce any of the C$_3$H$_2$ isomers \citep{Miller2003}, and we assume that this reaction does not occur at 10 K. Reactions with O and N atoms are one of the main destruction channels of CH$_2$CCH in the chemical model and the main cause of the low calculated abundance of this radical. If we assume that, as the C$_3$H$_4$ isomers, the propargyl radical does not react with O and N atoms, then the calculated abundance of CH$_2$CCH experiences an important enhancement, approaching those of the C$_3$H$_4$ isomers and becoming closer to the observed value (see black dashed line in Fig.~\ref{fig:abun}). A study of the reactions of CH$_2$CCH with O and N atoms at low temperatures would certainly shed light on the chemistry of this important radical. If these reactions are found to be rapid at low temperatures, then a powerful formation mechanism of propargyl, in addition to reactions (\ref{reac:c+c2h4}) and (\ref{reac:c3hn+}), would be needed to explain the high abundance of CH$_2$CCH observed in \mbox{TMC-1}. Alternatively, the gas-phase abundance of O atoms in \mbox{TMC-1} could be lower than given by the chemical model if oxygen is significantly depleted on grains. If the C/O gas-phase elemental ratio is assumed to be above one, then the calculated abundance of propargyl increases up to the observed value. However, the abundances of other C-bearing molecules increase also, in many cases well above the observed values (see, e.g., \citealt{Agundez2013}).

Reactions involving CH$_2$CCH can be an important source of organic molecules of a certain chemical complexity, including aromatic rings, under cold conditions. As a matter of fact, CH$_2$CCH is one of the most abundant radicals found in \mbox{TMC-1} and, given its radical nature, it is expected to show an enhanced reactivity compared to closed-shell molecules. On the other hand, CH$_2$CCH is a resonance-stabilized radical because it has a $\pi$-conjugated system (see discussion by \citealt{Tanaka1997}), which makes it to be less reactive than other radicals that do not have electron delocalization.

An illustration of a route to complex molecules under cold conditions is provided by the reactions of C$_3$H$_4$ isomers with CN and C$_2$H radicals \citep{Carty2001,Balucani2002,Zhang2009}, which are thought to form C$_4$H$_3$N and C$_5$H$_4$ isomers, respectively, in \mbox{TMC-1} \citep{Marcelino2021,Cernicharo2021d}. Similarly, reactions of the C$_3$H$_3$ radical with HCN and C$_2$H$_2$ could do the same job producing C$_4$H$_3$N and C$_5$H$_4$ isomers, respectively, with an even higher efficiency, given that C$_3$H$_3$/C$_3$H$_4$\,$\sim$\,1 and HCN/CN\,$>$\,1 (and probably C$_2$H$_2$/C$_2$H\,$>$\,1). However, the propargyl radical does not react with either C$_2$H$_2$ or C$_4$H$_2$ at low temperature due to the existence of activation barriers \citep{daSilva2017,Indarto2011}. Therefore, reactions of CH$_2$CCH with polyynes and cyanopolyynes are probably not efficient to form complex molecules in cold dark clouds. Reactions of CH$_2$CCH with abundant radicals like CN, C$_2$H, and C$_4$H could however be efficient in producing complex molecules if they are rapid at low temperature, given the high abundance of the reactants involved. Information on the kinetics of these reactions is however not available. Another route to form large molecules from the propargyl radical consists of condensation reactions with hydrocarbon ions of the type
\begin{subequations} \label{reac:c3h3+ions}
\begin{align}
\rm CH_2CCH + C_{\textit n}H_{\textit m}^+ & \rightarrow \rm C_{\textit n + 3}H_{\textit m + 2}^+ + H, \\
                                                                      & \rightarrow \rm C_{\textit n + 3}H_{\textit m + 1}^+ + H_2.
\end{align}
\end{subequations}
These reactions were assumed to proceed fast by \cite{Herbst1989}, although some empirical or theoretical evidence is needed to verify this hypothesis. In particular, it will be of great interest to investigate if these kind of reactions can lead to aromatic rings.

The CH$_2$CCH radical has a key role in the synthesis of benzene and polycyclic aromatic hydrocarbons (PAHs) in combustion processes, where it is thought that the reaction of two propargyl radicals is the main route to the formation of the first aromatic ring \citep{Miller1992}. The recent detection of benzonitrile (C$_6$H$_5$CN) in \mbox{TMC-1} \citep{McGuire2018} and the experimental evidence that it is most likely formed from benzene reacting with CN \citep{Cooke2020} highlights that benzene, and perhaps PAHs as well, are formed in situ in cold dark clouds. These environments have conditions very different to those of combustion flames, where the pressures and temperatures allow three-body recombination processes and reactions with activation barriers to proceed efficiently. However, the detection of abundant propargyl in \mbox{TMC-1} presented here suggests that this radical could play a key role in the cyclization towards the first aromatic ring, as it does in combustion chemistry.

\cite{Herbst1989} consider that the reaction
\begin{equation} \label{reac:c3h3+c3h3+}
\rm CH_2CCH + C_3H_3^+ \rightarrow \rm C_6H_5^+ + H
\end{equation}
proceeds fast resulting in the cyclic species C$_6$H$_5$$^+$, which can then lead to benzene after radiative association with H$_2$ and dissociative recombination of the ion C$_6$H$_7$$^+$ with electrons \citep{McEwan1999}. Reaction (\ref{reac:c3h3+c3h3+}) is included in our chemical model and it turns out to be one of the main sources of C$_6$H$_5$$^+$, and thus of benzene, in cold dark clouds. In the protoplanetary nebula CRL\,618 the synthesis of benzene, detected by \cite{Cernicharo2001}, rely as well on the precursor ion C$_6$H$_5$$^+$, although in that case it is mainly formed through the radiative association of C$_2$H$_2$ and C$_4$H$_3$$^+$ \citep{Woods2002}.

Formation of benzene could also occur through the reaction CH$_2$CCH + C$_3$H$_4$ $\rightarrow$ C$_6$H$_6$ + H, although kinetic modeling of combustion experiments point to the presence of activation barriers for the two isomers of C$_3$H$_4$ \citep{Faravelli2000}. It is unknown if reactions of CH$_2$CCH with other closed-shell hydrocarbons, like the recently discovered C$_4$H$_4$ and C$_5$H$_4$ isomers \citep{Cernicharo2021c,Cernicharo2021d}, could lead to cyclization without activation barrier. The self-reaction of propargyl radicals
\begin{equation} \label{reac:c3h3+c3h3}
\rm CH_2CCH + CH_2CCH \rightarrow \rm products
\end{equation}
is a very attractive candidate to form benzene in cold dark clouds. Reaction (\ref{reac:c3h3+c3h3}) has been widely studied for conditions relevant to combustion experiments, as it is believed to be the key cyclization step to benzene and PAHs in flames. Reaction (\ref{reac:c3h3+c3h3}) is relatively fast at room temperature, with a slight negative dependence on temperature, and a variety of products like benzene, fulvene ($c$-C$_5$H$_4$=CH$_2$), and C$_6$H$_5$ + H, with branching ratios depending on the temperature \citep{Miller2001,Shafir2003}. It would be of great interest to study the behavior of reaction (\ref{reac:c3h3+c3h3}) at low temperatures. In particular, it would be interesting to see if the channels leading directly to benzene through radiative association and to C$_6$H$_5$ + H are open at the low temperatures of cold dark cloud conditions.

\section{Conclusions}

We reported the detection of the propargyl radical (CH$_2$CCH) in the cold dark cloud \mbox{TMC-1}. The high column density derived, $8.7 \times 10^{13}$~cm$^{-2}$, is similar to that of the closed-shell counterpart CH$_3$CCH, making it one of the most abundant radicals detected in \mbox{TMC-1}. Due to its high abundance, it probably plays a key role in the synthesis of large organic molecules. In particular, it could be the key species in the cyclization to form aromatic rings like benzene in cold dark clouds.

\begin{acknowledgements}

We acknowledge the anonymous referee for a constructive report. We acknowledge funding support from Spanish MICIU through grants AYA2016-75066-C2-1-P, PID2019-106110GB-I00, and PID2019-107115GB-C21 and from the European Research Council (ERC Grant 610256: NANOCOSMOS). M.A. also acknowledges funding support from the Ram\'on y Cajal programme of Spanish MICIU (grant RyC-2014-16277).

\end{acknowledgements}

\appendix

\section{Updates with respect to {\small UMIST RATE12}}

Table~\ref{table:reactions} lists the chemical reactions involving CH$_2$CCH, CH$_3$CCH, and CH$_2$CCH$_2$ used in the chemical model, which either are not included in the gas-phase chemical network {\small RATE12} from the {\small UMIST} database \citep{McElroy2013} or are included with different rate coefficients. We have adopted rate coefficient expressions which are adequate, or at least reasonable, for a kinetic temperature of 10\,K. Some of the reactions in Table~\ref{table:reactions} are included in the {\small RATE12} network but with rate coefficients expressions that result in unrealistic values at 10\,K. This is the case of the reactions CH + C$_2$H$_4$ and C$_2$H + CH$_3$CHCH$_2$. In those cases in which there is information on product branching ratios, this information has been taken into account.

\begin{table*}
\tiny
\caption{Reactions involving CH$_2$CCH, CH$_3$CCH, and CH$_2$CCH$_2$ modified with respect to the {\small UMIST RATE12} chemical network.}
\label{table:reactions}
\centering
\begin{tabular}{lcccl}
\hline \hline
\multicolumn{1}{l}{Reaction} & \multicolumn{1}{c}{$\alpha$ (cm$^3$\,s$^{-1}$)} & \multicolumn{1}{c}{$\beta$} & \multicolumn{1}{c}{$\gamma$ (K)} & \multicolumn{1}{l}{Reference} \\
\hline
CH + C$_2$H$_4$ $\rightarrow$ CH$_3$CCH + H         & $1.12 \times 10^{-10}$ &  0 &     0 & Value at 23\,K \citep{Canosa1997}; branching ratio \citep{Goulay2009}. \\
CH + C$_2$H$_4$ $\rightarrow$ CH$_2$CCH$_2$ + H & $2.74 \times 10^{-10}$ &  0 &     0 & Value at 23\,K \citep{Canosa1997}; branching ratio \citep{Goulay2009}. \\
C$_2$H + CH$_3$CHCH$_2$ $\rightarrow$ CH$_2$CHCCH + CH$_3$ & $1.79 \times 10^{-10}$ &  0 &     0 & Value at 79\,K and branching ratio \citep{Bouwman2012}. \\
CH$_2$ + C$_2$H$_2$ $\rightarrow$ CH$_2$CCH + H & $2.00 \times 10^{-11}$ &  0 & 3330 & \cite{Baulch2005}. UMIST RATE12 incorrectly used expression for $^1$CH$_2$. \\
C$_2$ + CH$_4$ $\rightarrow$ CH$_2$CCH + H         & $1.30 \times 10^{-11}$ &  0 &     0 & Value at 24\,K \citep{Canosa2007}. \\
C + C$_2$H$_4$ $\rightarrow$ CH$_2$CCH + H         & $3.10 \times 10^{-10}$ &  0 &     0 & Value at 15\,K \citep{Chastaing1999}; branching ratio \citep{Bergeat2001}. \\
C$_2$H + CH$_3$ $\rightarrow$ CH$_2$CCH + H         & $1.00 \times 10^{-10}$ &  0 &     0 & \cite{Loison2017}. \\
C + CH$_3$CCH $\rightarrow$ C$_4$H$_3$ + H         & $2.47 \times 10^{-10}$ &  0 &     0 & Value at 298\,K \citep{Loison2004}. \\
C + CH$_3$CCH $\rightarrow$ C$_4$H$_2$ + H$_2$ & $4.35 \times 10^{-11}$ &  0 &     0 & Value at 298\,K \citep{Loison2004}. \\
C + CH$_2$CCH$_2$ $\rightarrow$ C$_4$H$_2$ + H$_2$ & $2.70 \times 10^{-10}$ &  0 &     0 & Value at 298\,K \citep{Loison2004}. \\
O + CH$_2$CCH $\rightarrow$ C$_2$H$_3$ + CO      & $2.30 \times 10^{-10}$ &  0 &     0 & Value at 295-750\,K \citep{Slagle1991}. The main likely product is HCCCHO \\
      & &  &     &  (see \citealt{Loison2017}), but it is not included in the network. \\
N + CH$_2$CCH $\rightarrow$ HC$_3$N + H$_2$      & $5.00 \times 10^{-11}$ &  0 &     0 & \cite{Loison2017}. \\
N + CH$_2$CCH $\rightarrow$ C$_2$H$_2$ + HCN   & $5.00 \times 10^{-11}$ &  0 &     0 & \cite{Loison2017}. \\
OH + CH$_2$CCH $\rightarrow$ C$_2$H$_3$ + HCO   & $6.00 \times 10^{-11}$ &  0 &     0 & \cite{Loison2017}. \\
OH + CH$_2$CCH $\rightarrow$ C$_2$H$_4$ + CO   & $6.00 \times 10^{-11}$ &  0 &     0 & \cite{Loison2017}. \\
 \hline
\end{tabular}
\tablenotea{\\
The rate coefficient is given by the expression $k(T) = \alpha (T/300)^\beta \exp(-\gamma/T)$, where $T$ is the gas kinetic temperature in units of K.
}
\end{table*}

\end{document}